\documentclass[conference]{IEEEtran}
\IEEEoverridecommandlockouts
\usepackage{cite}
\usepackage{amsmath,amssymb,amsfonts}
\usepackage{graphicx}
\usepackage{textcomp}
\usepackage{xcolor}
\usepackage{comment}
\usepackage{amsmath}
\usepackage{subfigure}
\usepackage{multirow}
\begin{document}

\title{A Noise-aware Enhancement Method for Underexposed Images\\
}

\author{\IEEEauthorblockN{Chien-Cheng Chien}
\IEEEauthorblockA{
\textit{Tokyo Metropolitan University}\\
Tokyo, Japan}
\and
\IEEEauthorblockN{Yuma Kinoshita}
\IEEEauthorblockA{
\textit{Tokyo Metropolitan University}\\
Tokyo, Japan }
\and
\IEEEauthorblockN{Hitoshi Kiya}
\IEEEauthorblockA{
\textit{Tokyo Metropolitan University}\\
Tokyo, Japan}
}

\maketitle

\begin{abstract}
A novel method of contrast enhancement is proposed for underexposed images, in which heavy noise is hidden.
Under low light conditions, images taken by digital cameras have low contrast in dark or bright regions.
This is due to a limited dynamic range that imaging sensors have.
For these reasons, various contrast enhancement methods have been proposed so far.
These methods, however, have two problems: (1) The loss of details in bright regions due to over-enhancement of contrast. (2) The noise is amplified in dark regions because conventional enhancement methods do not consider noise included in images.
The proposed method aims to overcome these problems.
In the proposed method, a shadow-up function is applied to adaptive gamma correction with weighting distribution, and a denoising filter is also used to avoid noise being amplified in dark regions.
As a result, the proposed method allows us not only to enhance
contrast of dark regions, but also to avoid amplifying noise, even under strong noise environments.
\end{abstract}


\section{Introduction}
To overcome a limited dynamic range that imaging sensors have, various contrast enhancement methods have so far been proposed.
The histogram equalization (HE) is one of the most popular algorithms for contrast enhancement, and there are various extended versions of the HE.
However, these histogram-based methods cause the loss of details in bright regions due to the over-enhancement.
Contrast enhancement methods based on the Retinex theory have also been studied\cite{WVM,LIME,EDW}.
Although these methods can enhance the contrast while preserving details in bright areas, they also have a noise amplification problem as with histogram-based methods.

To avoid the noise amplification problem, some contrast enhancement methods have been proposed\cite{WVM,SUH,LIME}.
However, they do not preserve details in bright areas, although, they can reduce some noise.

Because of such a situation, we proposes a novel image contrast enhancement method based on both the Retinex theory and a noise aware shadow-up function.
The proposed method can enhance image contrast without over-enhancement and noise amplification.
A shadow-up function is used for preventing over-enhancement and the loss of details in bright regions.
In addition, the use of a mapping function designed by using adaptive gamma correction with weighting distribution (AGCWD) \cite{AGCWD} allows not only to enhance contrast in dark regions, but also to avoid amplifying noise.

In an experiment, the proposed method is compared with conventional contrast enhancement methods, including state-of-the-art ones.
Experimental results show that the proposed method can produce high quality images without over-enhancement and noise amplification.
%
\begin{figure*}
\begin{center}
\begin{tabular}{cc}
\begin{minipage}[]{.64\textwidth}
\centering
\includegraphics[height=50mm]{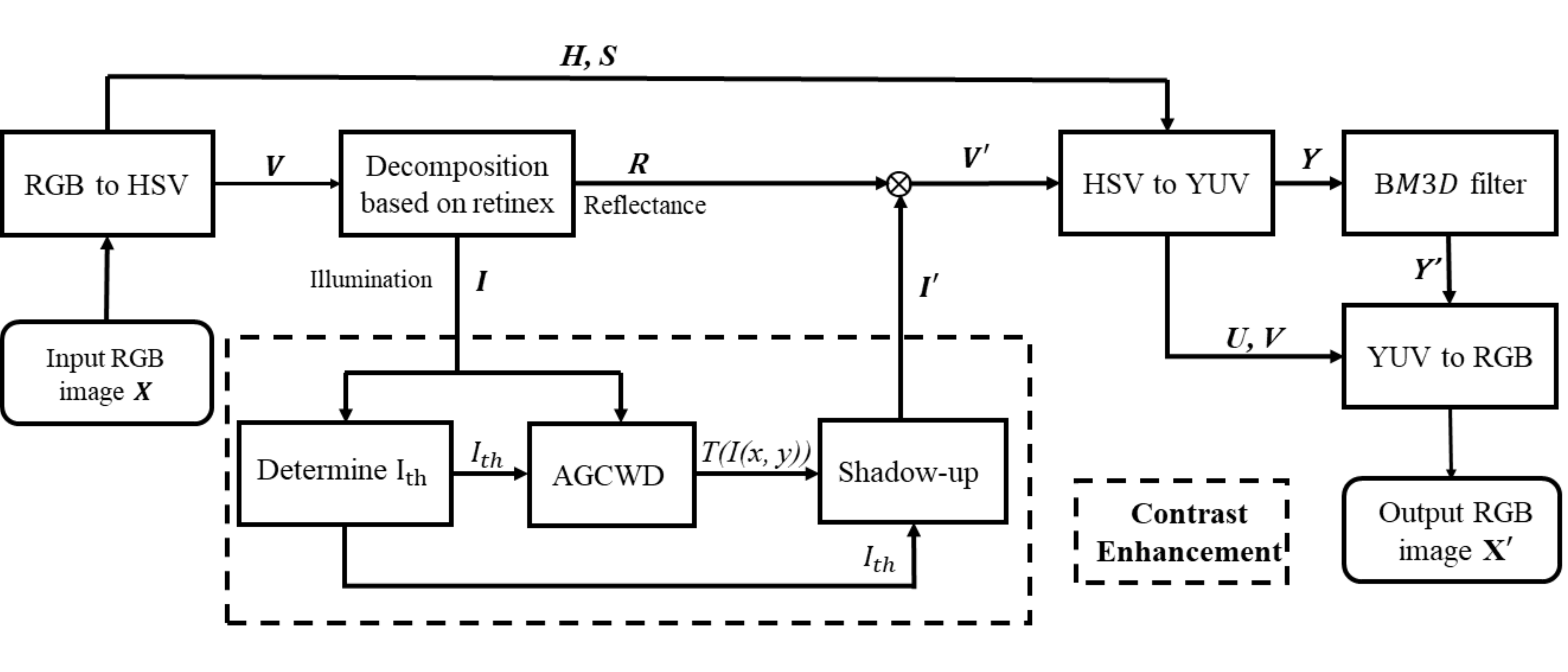}
\caption{Flowchart of the proposed method.}
\label{fig:flowchart} 
\end{minipage}&
\begin{minipage}[]{.35\textwidth}
\centering
\includegraphics[height=50mm]{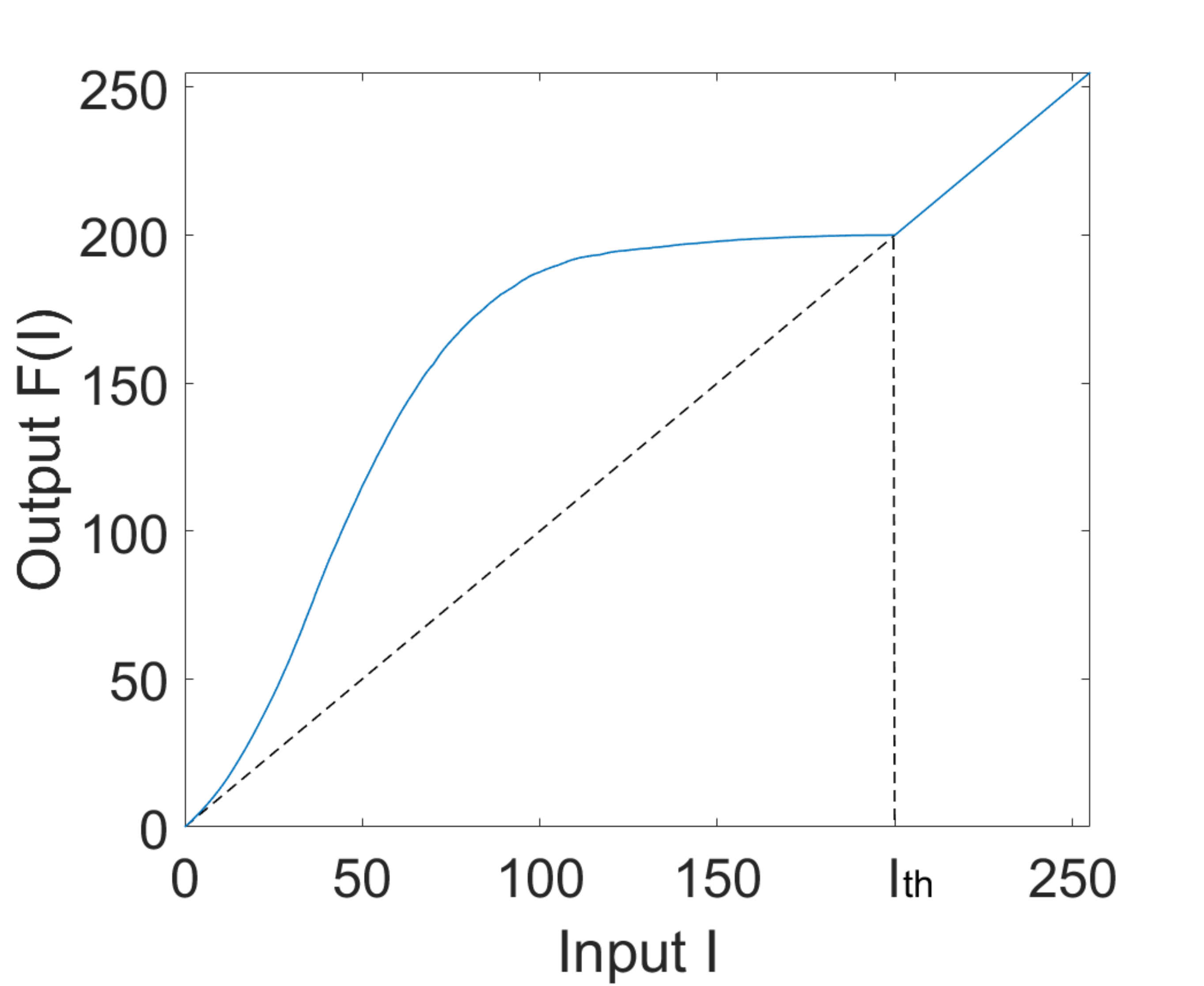}
\caption{Example of mapping curves.}
\label{fig:shadow-up}
\end{minipage}
\end{tabular}
\end{center}
\end{figure*} 
\section{RELATED WORKS}
Related works are summarized here.
\subsection{Retinex theory}
\label{sec:Retinex}
Retinex theory is based on the relation, $S=R \cdot L$, where original image S is the product of illumination L and reflectance R.
When the information of only one surround is used for the conversion of each pixel, its approach is called Single-Scale Retinex (SSR)\cite{SSR}.
In SSR, halo artifacts occur unnaturally in the boundary of regions with large gradient values.
To solve this problem, Multi-Scale Retinex (MSR)\cite{MSR} was proposed.
However, since a logarithmic transformation is used, MSR still causes a problem that the results do not stabilize due to the influence of noise in dark areas. 
Simultaneous reflection \& illumination estimation (SRIE)\cite{SRIE} and weighted variation model (WVM)\cite{WVM} are also Retinex-based methods.
These methods have a good performance for images without noise, but some strange areas are generated in strong noise environments.
Therefore, many outstanding methods\cite{DJJ,LIME,XRE} have been proposed to improve the quality of images, and preserve more details.
\subsection{Image enhancement}
 \label{sec:exposure}
The histogram equalization (HE) \cite{SPI} is one of the most popular algorithms for contrast enhancement\cite{HKS} and various  extended versions of HE have been proposed \cite{KZU,YK,YW,AGCWD,XWU, chien2019retinex}. 
Contrast enhancement using adaptive gamma correction with weighting distribution (AGCWD)\cite{AGCWD} aims to prevent over-enhancement and under-enhancement caused by using adaptive gamma correction and a modified probability distribution. 
However, the over-enhancement and the loss of contrast in bright areas are still caused under the use of these histogram-based methods.
Some noise hidden in the darkness is also amplified. Because of such a situation, a number of histogram-based contrast enhancement methods have been proposed to prevent the noise amplification.
In the methods, a shrinkage function is used for preventing the noise amplification.
Low light image enhancement based on two-step noise suppression (LLIE)\cite{SUH} uses both noise level function (NLF) and just noticeable difference (JND) for contrast enhancement with noise suppression.
Although this method can reduce some noise, it does not preserve details in bright areas as with histogram-based methods.
Another way for enhancing images is to use a multi-exposure image fusion method by using photos with different exposures \cite{MEI, AEC, kinoshita2018automatic_trans, kinoshita2019scene}.
\subsection{Deoising filter}
\label{sec:deoising}
Image denoising has a great tradition in the research field of signal processing because of its fundamental role in many applications. 
In particular, block-matching and 3D filtering (BM3D) \cite{KDA} is one of the most successful advances.
In this paper, BM3D is used as one of noise suppression techniques.
Our purpose is not only to enhance contrast with noise suppression, but also to preserve details in bright regions based on Retinex theory.
%
\begin{figure*}[!t]
\begin{minipage}[]{.99\textwidth}
\centering
\subfigure[Original image] {
\label{subfigure:10}
\includegraphics[width=.225\textwidth]{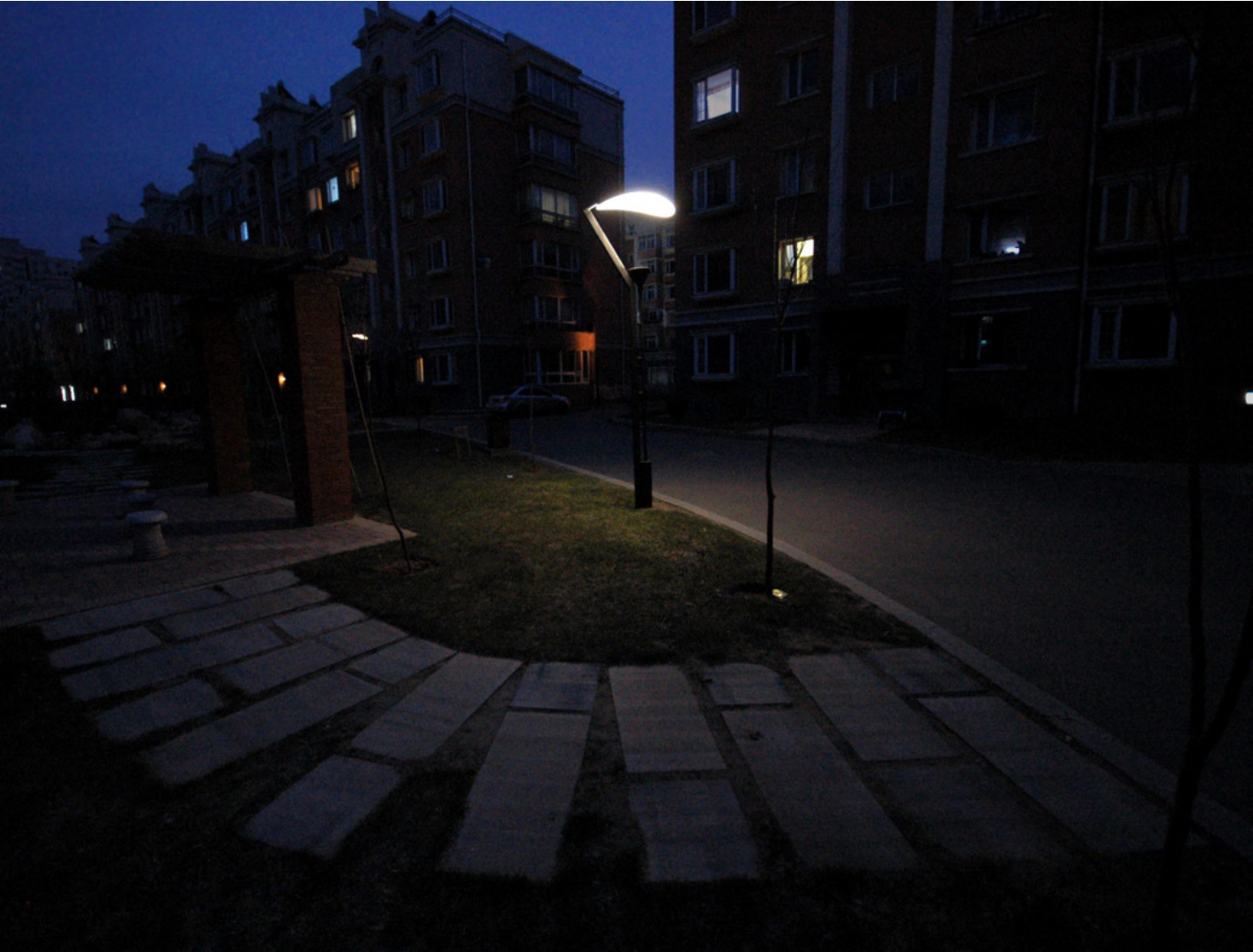}
}
\hspace{-4mm}
\subfigure[Reflectance layer] {
\label{subfigure:10_1}
\includegraphics[width=.225\textwidth]{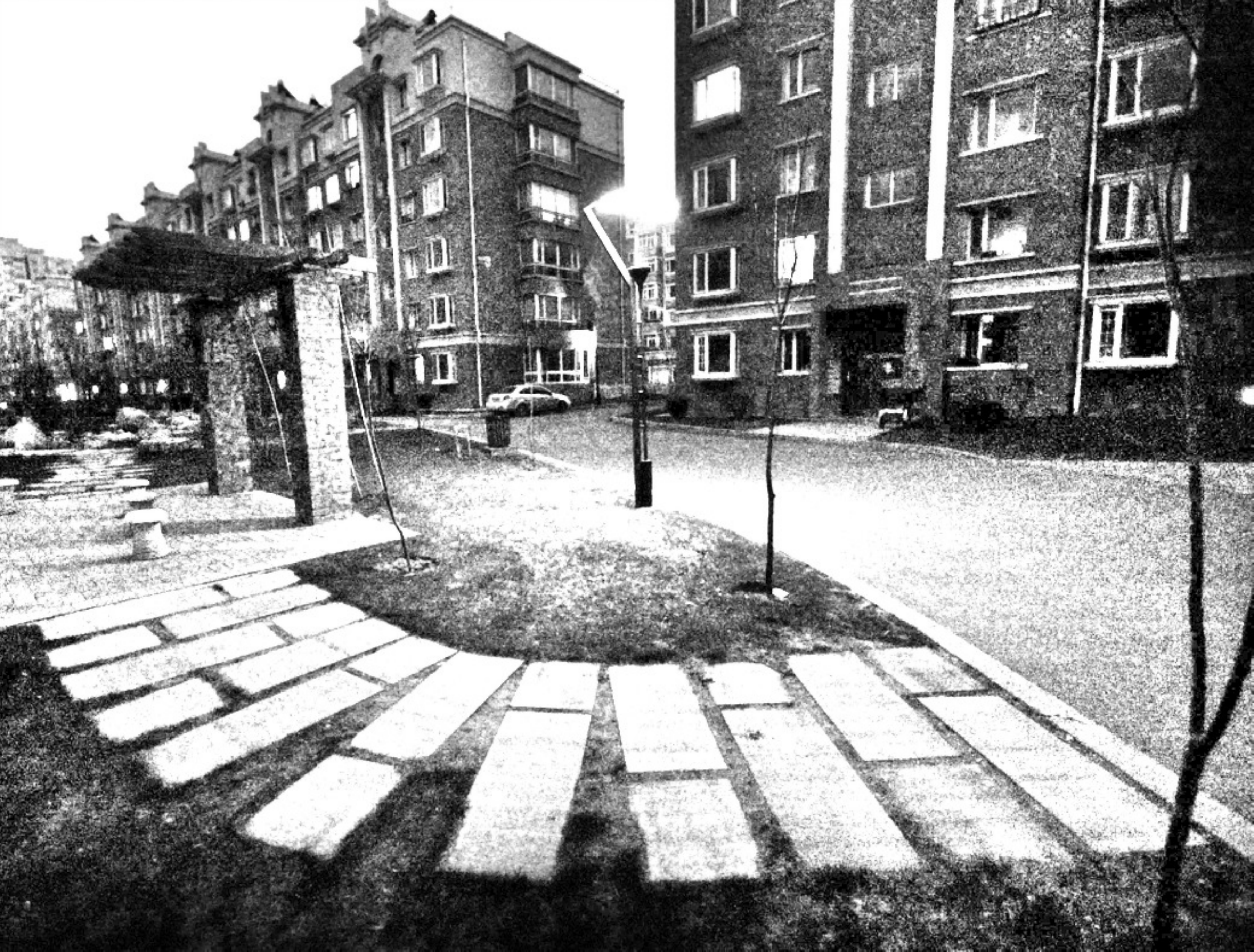}
}
\hspace{-4mm}
\subfigure[ Illumination layer] {
\label{subfigure:10_2}
\includegraphics[width=.225\textwidth]{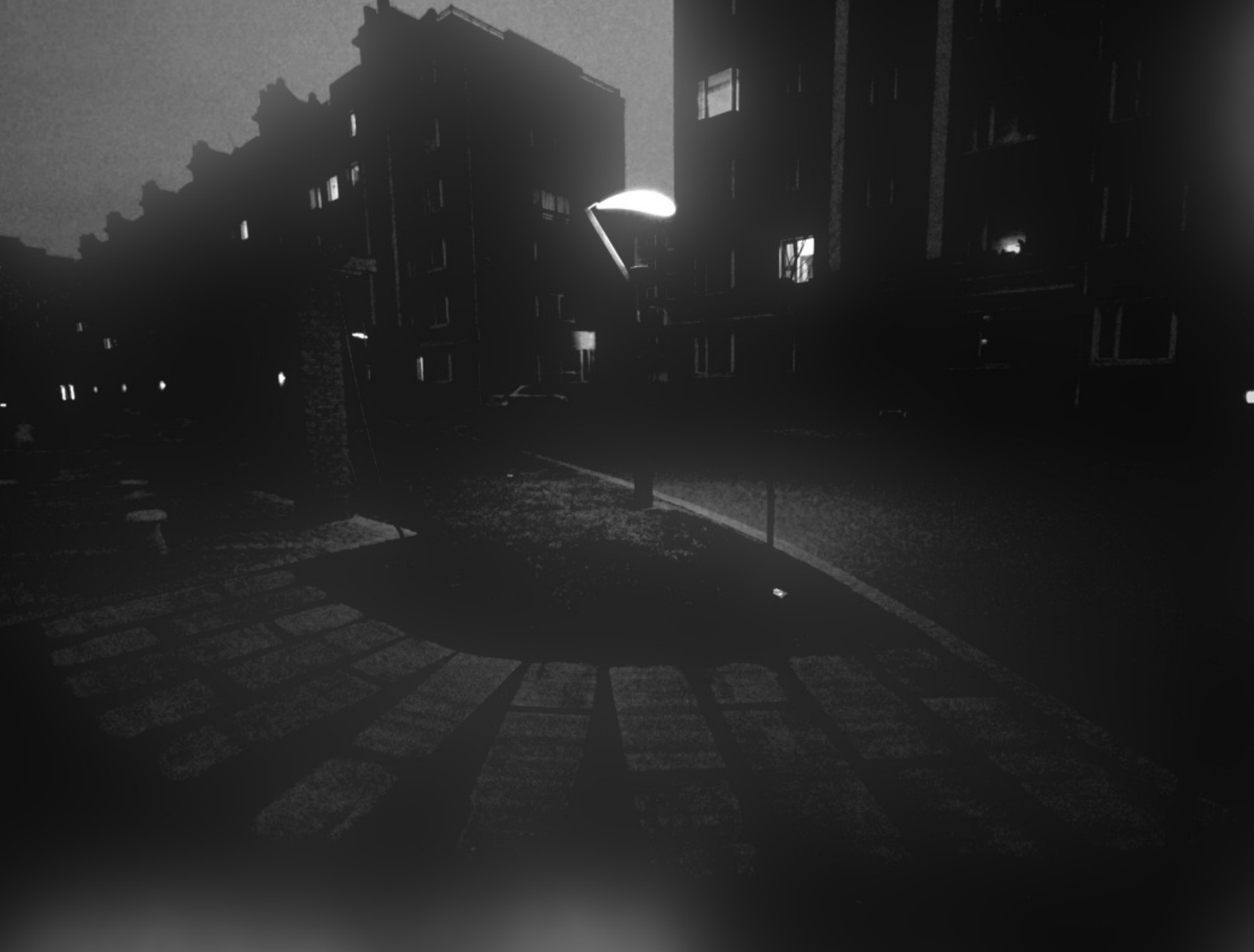}
}
\hspace{-4mm}
\subfigure[Enhanced  illumination layer] {
\label{subfigure:10_3}
\includegraphics[width=.225\textwidth]{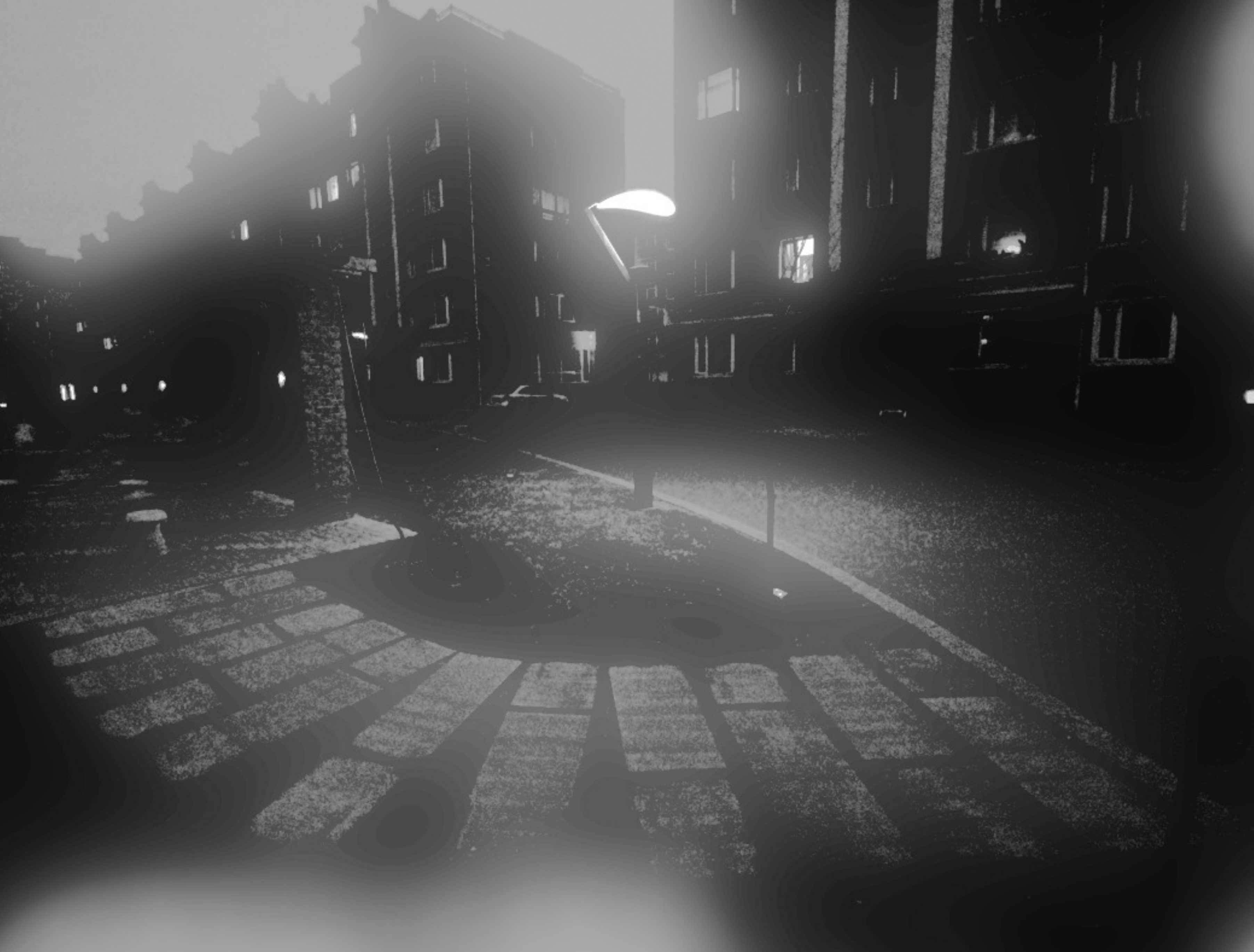}
}
\caption{Example of weighted variation model (WVM), and example of illumination layer is enhanced by AGCWD with shadow-up function.}
\label{fig:3}
\end{minipage}
\end {figure*}
\begin{figure*}[!t]
\begin{minipage}[]{.99\textwidth}
\centering
\subfigure[Original image] {
\label{subfigure:orig}
\includegraphics[width=.225\textwidth]{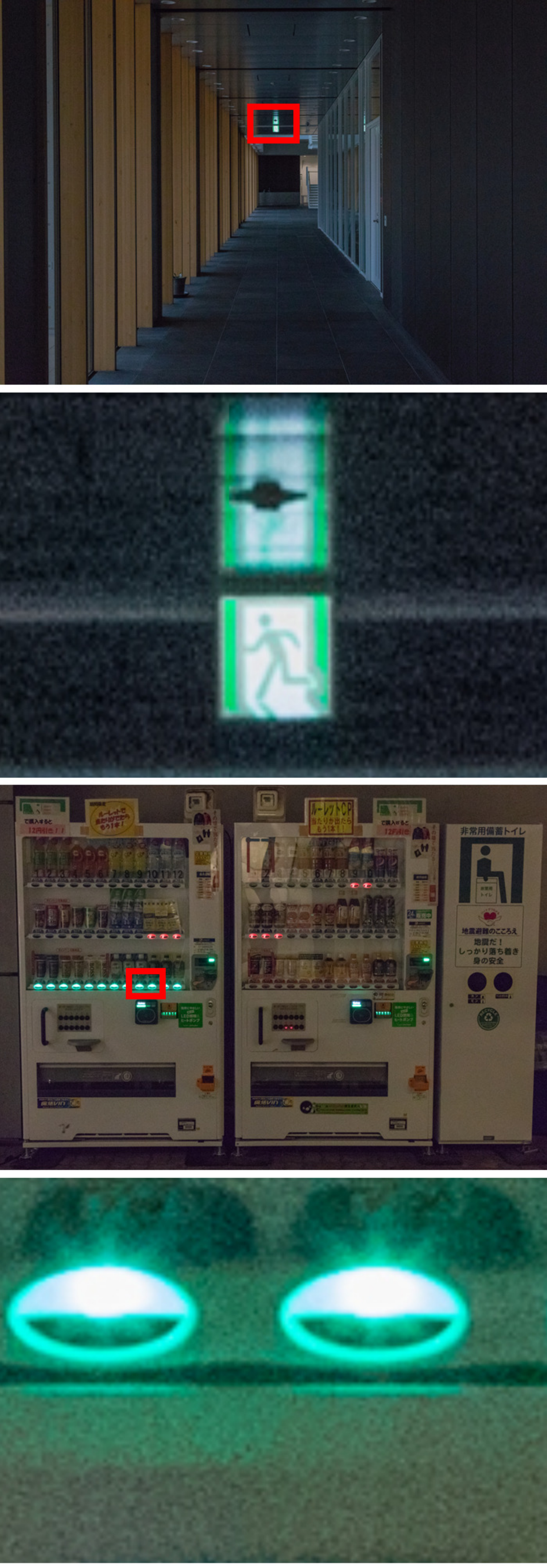}
}
\hspace{-4mm}
\subfigure[AGCWD\cite{AGCWD}] {
\label{subfigure:agc}
\includegraphics[width=.225\textwidth]{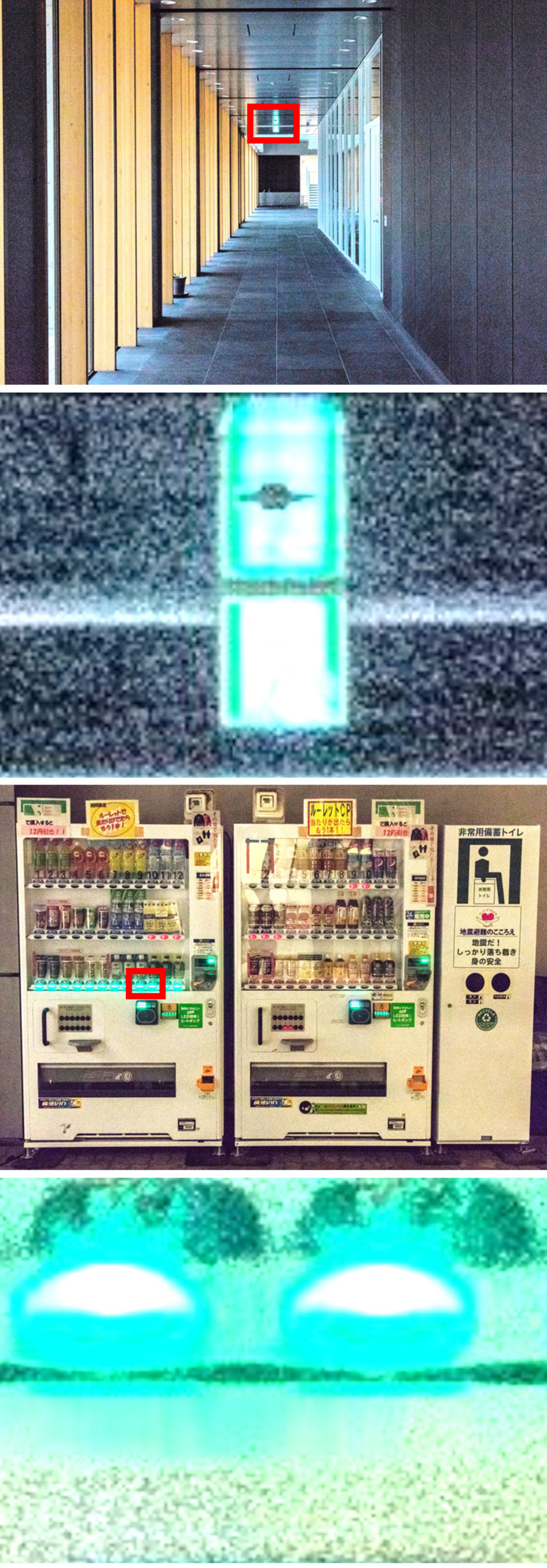}
}
\hspace{-4mm}
\subfigure[WVM\cite{WVM}] {
\label{subfigure:wvm}
\includegraphics[width=.225\textwidth]{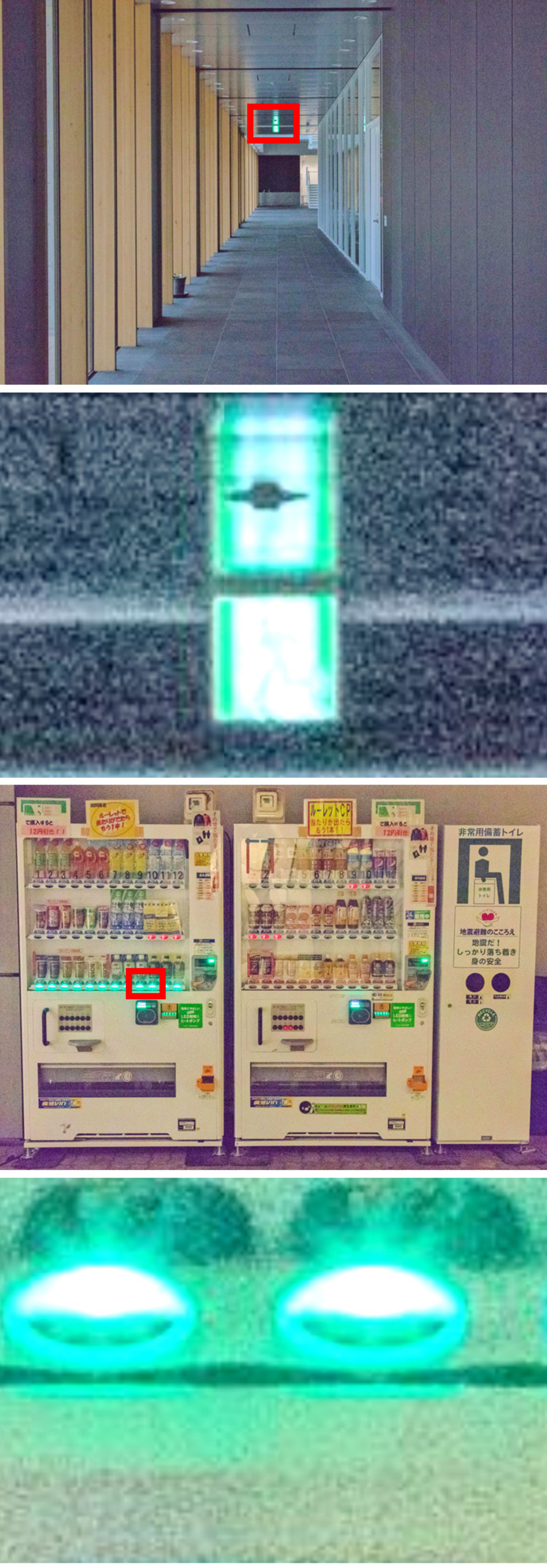}
}
\hspace{-4mm}
\subfigure[Proposed method] {
\label{subfigure:pro}
\includegraphics[width=.225\textwidth]{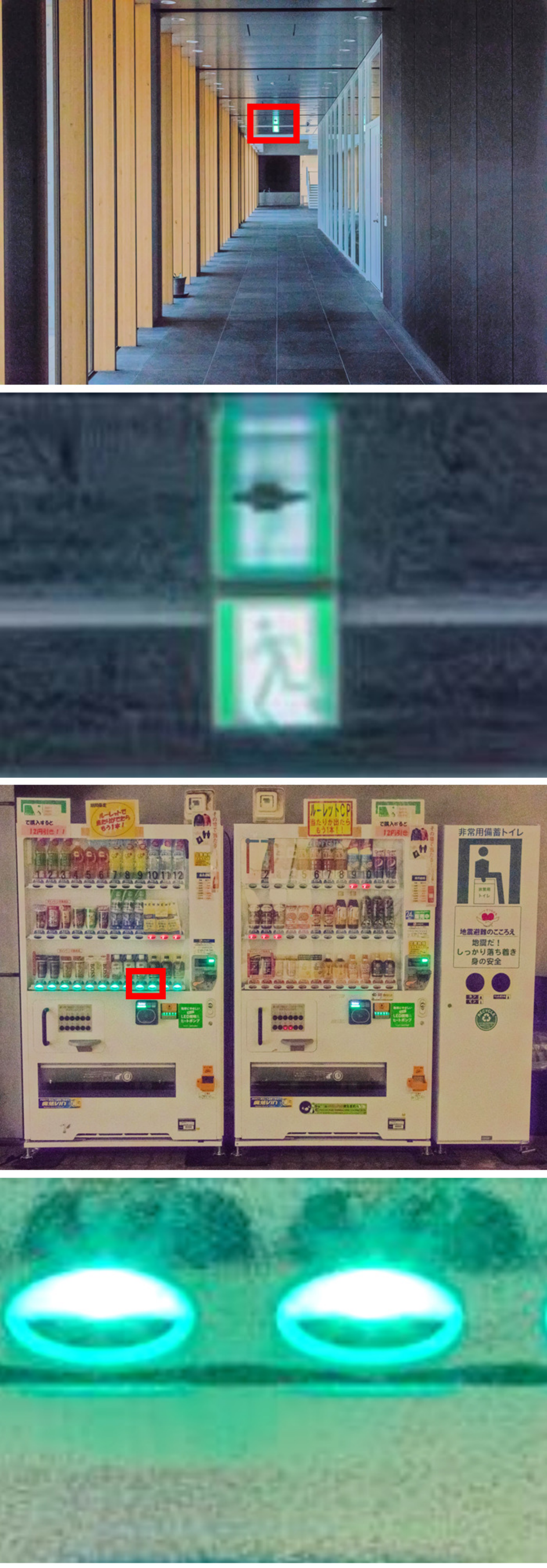}
}
\caption{Experimental Results.}
\label{fig:4}
\end{minipage}
\end {figure*}
%
\section{Proposed method}
The novelty and the detail of the proposed method are explained here, and the outline of the proposed method is shown in Fig \ref{fig:flowchart}.
\subsection{Decomposition based on Retinex}

As shown in Fig. \ref{fig:flowchart}, an input RGB image $\boldsymbol{X}=\{\boldsymbol{X}_{R},\boldsymbol{X}_{G},\boldsymbol{X}_{B}\}$ is transformed to an HSV image $\boldsymbol{X}_{HSV}=\{\boldsymbol{H},\boldsymbol{S},\boldsymbol{V}\}$, where $\boldsymbol{H}$, $\boldsymbol{S}$ and $\boldsymbol{V}$ are hue, saturation and brightness images, respectively.
An excellent weighted variational model (WVM) was proposed for simultaneous reflectance and illumination estimation\cite{WVM}.
We use this model for decomposing $\boldsymbol{V}$ into illumination layer $\boldsymbol{I}$ and reflectance layer $\boldsymbol{R}$, where $\boldsymbol{R}$ has almost no noise, but $\boldsymbol{I}$ includes, due to the work of the model.

\subsection{Contrast Enhancement}

$\boldsymbol{I}$ is enhanced by using two key technologies: Shadow-up function and AGCWD.
The use of a shadow-up function aims to avoid the loss of details in bright areas due to over enhancement, and example is shown in Fig. \ref{fig:shadow-up}.  
A shadow-up function, which consists of a nonlinear part and a linear part, is given by 
\begin{equation}
\setlength{\abovedisplayskip}{3pt}
\setlength{\belowdisplayskip}{3pt}
I^\prime(x,y) = \left\{
		\begin{aligned}
                 &T(I(x,y))&&\mbox{, if $I(x,y)<I_{th}$} \\    
                 &I(x,y)&&\mbox{, otherwise} \\  
		\end{aligned}
		\right.
		,
\label{eq1}
\end{equation}
where $I(x,y)\in[0, 255]$ is the intensity of illumination layer at a coordinate $(x,y)$, $T(I(x,y))$ is a monotonically increasing function, and $I_{th}$ is an upper limit of the nonlinear part for avoiding over enhancement in bright areas.
Contrast is enhanced only when $I(x,y)$ is less than the threshold value $I_{th}$, according to (1). 

To determine a proper threshold value $I_{th}$ for illumination layer, we take into account the luminance distribution of the illumination layer.
Let it be $H=\{(x,y):I_{th}<I(x,y)<I_{max}\}$, where $I_{th}$ is the $th$ percentile of luminance $I(x,y)$ of the input image, and $I_{max}$ is the maximum of $I(x,y)$.
The threshold value $I_{th}$ is calculated as follows:
\begin{equation}
\setlength{\abovedisplayskip}{3pt}
\setlength{\belowdisplayskip}{3pt}
I_{th}=255-\frac{1}{\arrowvert H \arrowvert}\sum_{(x,y)\in{H}}I(x,y).
\label{eq2}
\end{equation}
A threshold value $I_{th}$  for a bright image becomes smaller than for a darker image.

AGCWD is a method to design a function $T(I(x,y))$.
However, AGCWD usually causes a noise amplification problem, because it does not consider the influence of noise included in images\cite{SUH}. 
To overcome this problem,  both the Retinex theory and $I_{th}$ are applied to AGCWD in this paper.
From Fig. \ref{fig:3}, Fig. \ref{subfigure:10_1} and Fig. \ref{subfigure:10_2} are reflectance layer and illumination layer based on Retinex theory, and Fig. \ref{subfigure:10_2} is a illumination layer which is enhanced by AGCWD with shadow-up function.
\subsection{Output YUV image}

By enhancing the illumination layer $\boldsymbol{I}$, an adjusted illumination $\boldsymbol{I^{\prime}}$ is obtained. Then an enhanced  $\boldsymbol{V^{\prime}}$ is computed by $V^{\prime}(x,y)=I^{\prime}(x,y)\cdot R(x,y)$.
Finally, an YUV image $\boldsymbol{Y}_{YUV}=\{\boldsymbol{Y},\boldsymbol{U},\boldsymbol{V}^\prime\}$ is obtained by using $\boldsymbol{V^{\prime}}, \boldsymbol{H}$, and $\boldsymbol{S}$, according to the model\cite{WVM}.

\subsection{Denoising technique}

$\boldsymbol{Y}_{YUV}$ still has some noise in dark areas, since $\boldsymbol{R}$ includes noise, though the Retinex theory allows us to avoid enhancing the noise.
Therefore, a denoising technique is required to further improve the visual quality.
Block-matching and 3D filtering (BM3D) \cite{KDA} is chosen as a denoising method in this paper. 
In our implementation, for further cutting the computational load, BM3D is applied to only $\boldsymbol{Y}$ channel.
After the denoising, an enhanced RGB image $\boldsymbol{X}^\prime$ is computed by using $\boldsymbol{Y}^\prime$, $\boldsymbol{U}$ and $\boldsymbol{V}$.

\section{Simulation}
\subsection{Simulation condition}
We used six images for our simulation, where four images were from LIME \cite{LIME}, and two other images were taken by a digital camera Canon 5D mark IV under the conditions: ISO 12800 and safe shutter speed.
Because the ISO value was very high, the two images contained a lot of noise as shown in Fig. \ref{subfigure:orig}.
We carried out a simulation to compare the proposed method with conventional contrast enhancement methods, AGCWD\cite{AGCWD} and WVM\cite{WVM} and LIME\cite{LIME}.
We adopted the $75th$ percentile as $I_{th}$.
\subsection{Simulation results}
\subsubsection{Visual comparison} 
We picked up two images taken by the camera resulting images, subjectively in Fig. \ref{fig:4}.
The second columns in Fig. \ref{fig:4} are the enlarged view of red boxes in the first column, so that we clearly see the difference in dark and bright areas, and noise.
From Fig. \ref{subfigure:agc}, it is confirmed that AGCWD over-enhanced bright areas, but clearly enhanced dark areas. 
Further, we easily see a lot of noise in the image. 
WVM not only enhanced noise but also changed the white balance in dark areas. Also, we easily observe unusual purple areas in Fig. \ref{subfigure:wvm}. 
In contrast, the proposed method provided almost same quality as that of the original image in bright areas.
Moreover, the image had less noise than AGCWD and WVM in dark areas.
\subsubsection{Objective evaluation} 
A blind image quality assessments, called natural image quality evaluator (NIQE)\cite{NIQE}
was used to objectively evaluate the quality of enhanced images.
Here, a matlab function \textit{niqe()} and its default model were used for the evaluation.
Since the default model used in \textit{niqe()} were trained with noisy images, 
a lower NIQE score represents that the evaluated image has less noise.

Table \ref{table:1} shows NIQE scores for six images.
From the table,
we can confirm that the proposed method averagely had lower scores
than the other methods including the state-of-the-art ones.
Hence, the proposed method was demonstrated to be
effective to avoid noise amplification while enhancing image contrast.
\begin{table}[t]
\caption{EXPERIMENTAL RESULTS (NIQE)}
\label{table:1}
\centering
\begin{tabular}{c|cccc|c}
\hline
\hline
Method&Original&AGCWD\cite{AGCWD}&WVM\cite{WVM}&LIME\cite{LIME}&Proposed\\ 
\hline
Image 1 & \textbf{4.795} & 5.327          & 5.054          & 5.087  & 6.045          \\
Image 2 & 5.692          & 7.712          & 7.792          & 10.676 & \textbf{5.094} \\
Image 3 & 4.366          & 4.169          & \textbf{3.754} & 4.755  & 5.245          \\
Image 4 & 8.963          & 9.260          & 9.757          & 10.656 & \textbf{3.809} \\
Image 5 & 6.497          & 6.807          & 6.718          & 7.487  & \textbf{4.257} \\
Image 6 & 3.226          & \textbf{3.196} & 3.221          & 3.777  & 4.344          \\ \hline
Average & 5.590          & 6.078          & 6.050          & 7.073  & \textbf{4.799} \\ \hline
\hline
\end{tabular}
\end{table}
%
\section{Conclusion}
We proposed a novel image contrast enhancement method based on both the Retinex theory and a noise aware shadow-up function.
The proposed method can enhance the contrast of images without over-enhancement and noise amplification.
Experimental results showed that the proposed method successfully enhances contrast, while preserving details in bright regions and suppressing some noise in dark regions.

\end{document}